# Skill requirements in job advertisements: A comparison of skill-categorization methods based on explanatory power in wage regressions[1]

Ziqiao Ao[2]; Gergely Horvath[3]; Chunyuan Sheng[4]; Yifan Song[5]; Yutong Sun[6]

## Abstract

*In this paper, we compare different methods to extract skill requirements from job advertisements. We consider three top-down methods that are based on expert-created dictionaries of keywords, and a bottom-up method of unsupervised topic modeling, the Latent Dirichlet Allocation (LDA) model. We measure the skill requirements based on these methods using a U.K. dataset of job advertisements that contains over 1 million entries. We estimate the returns of the identified skills using wage regressions. Finally, we compare the different methods by the wage variation they can explain, assuming that better-identified skills will explain a higher fraction of the wage variation in the labor market. We find that the top-down methods perform worse than the LDA model, as they can explain only about 20% of the wage variation, while the LDA model explains about 45% of it.*

---

[1] We acknowledge the financial support of the Office of Undergraduate Studies at the Duke Kunshan University under the Summer Research Scholars program. Authors are listed in alphabetical order.

[2] McCormick School of Engineering, Northwestern University, Address: 633 Clark St, Evanston, IL 60208, USA, E-mail: ZiqiaoAo2023@u.northwestern.edu

[3] *Corresponding author.* Division of Social Sciences, Duke Kunshan University. Address: Division of Social Sciences, Duke Kunshan University, No. 8 Duke Avenue, Kunshan, Jiangsu Province, China 215316. E-mail address: horvathgergely@gmail.com . Tel: +86 155-01683821.

[4] Graduate Institute Geneva, Address: Chem. Eugène-Rigot 2, 1202 Genève, Switzerland, E-mail: chunyuan.sheng@dukekunshan.edu.cn

[5] Division of Natural Sciences, Duke Kunshan University, Address: Division of Social Sciences, Duke Kunshan University, No. 8 Duke Avenue, Kunshan, Jiangsu Province, China 215316. E-mail: yifan.song348@dukekunshan.edu.cn

[6] Division of Natural Sciences, Duke Kunshan University, Address: Division of Social Sciences, Duke Kunshan University, No. 8 Duke Avenue, Kunshan, Jiangsu Province, China 215316. E-mail: yutong.sun@dukekunshan.edu.cn

# 1. Introduction

Job advertisements on online job boards provide important and real-time information on the state of the labor market. They indicate the temporal fluctuations and geographical variation in labor demand, as well, as the structure of labor demand in terms of occupations and industries. Job board data has also been used to come up with new classification methods of jobs to occupations (see e.g., Djumalieva et al., 2018; Séguéla and Saporta, 2010) extract skill requirements for specific professions (see e.g., Ham et al., 2020; Jiang and Chen, 2021), to understand the wage premium paid for skills or to understand the dynamics and implications of skill-biased technological change (see e.g., Deming and Kahn, 2018; Hershbein and Kahn, 2018).

While the usefulness of job board data is evident, it is much less clear what methods researchers should use to classify jobs and extract skill requirements from job advertisements. In this paper, we aim to compare different methods to identify skill requirements in the text of job advertisements. On the one hand, we consider top-down, supervised methods that are based on a dictionary of keywords which were established by expert opinion. These methods are often used in the economics literature to investigate important questions in labor economics (see e.g., Atalay et al., 2020; Deming and Kahn, 2018; Hershbein and Kahn, 2018; Spitz-Oener; 2006). They usually establish a smaller number of skill categories that are general in the sense that they capture skills that may be demanded in many different types of jobs. For example, communication skills, computers skills, managerial skills, etc. Based on the established set of keywords, it is relatively straightforward to identify and count the skill-related keywords in the text of job advertisements. We consider the 5-skills categorization from Spitz-Oener (2006), the 10-skills categorization established by Deming and Kahn (2018), and the skill categorization of the European Dictionary of Skills and Competences (DISCO). The latter is a multilingual, peer-reviewed thesaurus used to classify, describe and translate skills and competencies, which has been incorporated in the European Classification of Skills, Competences, Occupations, and Qualifications. It is available on the website http://disco-tools.eu/disco2_portal/. DISCO categorizes skills into 9 non-domain specific, general skills and 25 domain-specific skills that are particular to different types of occupations.

We compare these top-down skill-categorization methods with a bottom-up, unsupervised topic modeling technique, the Latent Dirichlet Allocation (Blei et al., 2003). This method identifies the underlying topics in the job advertisement texts using the probability distribution of words in the texts, without further outside input. We interpret each topic as a specific skill requirement. LDA returns a set of keywords associated to each topic and also weights that measure how much a topic is represented in a given job advertisement. We use these weights to judge the skill requirements of a given job advertisement. We determine the number of topics/skill categories by maximizing the coherence score of the allocation suggested by the algorithm (as it is explained below in more detail), which gives 24 topics.

Besides the comparison of these methods coming from different disciplines, e.g., economics and computer science, the main contribution of our paper is to propose wage regressions as a method of comparison. More specifically, we compare the skill categorization methods by the amount of wage variation they can explain in the data. The underlying idea is that firms reward skills by higher wages and they set the wages paid by each job based on the quantity and type of skills the employee needs to possess to perform well in the job. Our dataset contains the wages paid by each job and the skill requirements are embedded in the job descriptions. A better skill categorization method thus should be able to perform better at extracting the relevant skill requirements from the job description and the more accurate the established skill requirements are, the higher fraction of the wage variation between jobs they can explain. To obtain a measure of the wage variation explained by skill requirements, we run wage regressions and compute the adjusted R2 from the regressions.

We perform this exercise on a job advertisement dataset from the Reed website in the UK, collected in 2018. Our dataset contains over 1 million job ads, and covers all types of jobs that are represented on the website.

Our results show the distribution of skill requirements over the job advertisements according to the different skill categorization methods. As for the wage regressions, we identify the different types of skills that pay a wage premium or discount. Most of our identified skill categories will have a statistically significant correlation with the wage. Regarding the comparison of different methods, we obtain that each of the three top-down skill categorization methods can explain 20% of the wage variation in the data, we do not find large differences between them. In contrast, the skill categorization based on the LDA topic modeling technique can explain over 40% of the wage variation. We thus find that the bottom-up, unsupervised method is superior to the top-down, supervised method in capturing the skill requirements embedded in the job advertisements. We also note, however, that one disadvantage of the LDA method is that the identified topic skills are harder to interpret than the clear-cut categorization based on expert opinion.

Our paper is organized as follows. In Section 2, we summarize the relevant literature from computer/data science and economics. In Section 3, we introduce our dataset and data cleaning methods. In Section 4, we describe the different skill categorization methods and the distribution of skill requirements in the data according to each categorization method. In Section 5, we present the wage regressions, the skill premia/discounts associated to the identified skills and compare the different methods based on the wage variation they can explain. Section 6 concludes the paper.

# 2. Literature Review

We review the relevant literature in two steps. Firstly, we review the data analytics and computer science literature which focused on the methodology of extracting skill requirements from job advertisements. Secondly, we focus on the economics literature which applied text analytics methods to answer important questions in labor economics.

## 2.1 Data science literature

The development of text analytics techniques and online job board websites made it possible to track the current changes in labor demand in a timely manner that was impossible before by other data sources. Job advertisements extracted from job board websites can be used to categorize jobs into occupations, understand the changes in skill requirements and the temporal fluctuations and geographical variations of labor demand.

A large applied literature studies the skill requirements of different occupations and their changes over time. For example, Ham et al. (2020) studies the skill requirements of auditors, Jiang and Chen (2021) analyzes the changes in demand for data science skills, Gao et al. (2020) investigates the demand for financial skills in the U.S. labor market, while Meyer (2019) studies the skill requirements of health care data scientists. The common feature of these studies is that they focus on a certain occupation where it is relatively easy to establish a smaller set of relevant skills that are characteristics of that occupation. Then simple search for the respective words in the job advertisements can be used to detect demand for the skills.

Another literature aimed to develop a classification of jobs into occupations and as such worked with more general datasets that cover a wide range of jobs. There have been different machine learning methodologies used by this literature, including supervised and unsupervised algorithms. For example, Djumalieva et al.

(2018) provide one of the first data-driven methodologies for grouping job advertisements into occupations based on the skills contained within those advertisements. They use skills mentioned in a job advert to understand the nature of the job, and they categorize jobs using supervised methods like the pre-trained word embeddings model, which compares the skills in each job advert to the existing DISCO job categorizations and decides which occupational category the job belongs to based on the similarities of the terms. Djumalieva et al. (2018) also apply unsupervised categorization methods, such as k-means clustering, to categorize jobs and find that the skill categories with the largest proportion of job adverts are sales and distribution. Another example is Séguéla and Saporta (2010) who construct a hierarchical categorization of jobs and propose a method to categorize the job descriptions into a two-level predefined classification system using a Support Vector Machine model. They apply their method to a corpus of 700 job postings from France.

Over time the literature has developed and applied several approaches to categorize jobs based on job advertisements and many researchers aimed to compare these categorization methods to find the one with the best performance. For example, Ikudo et al. (2019) compare the categorization methods among Multinomial Naïve Bayes, Bernoulli Naïve Bayes, Random Forests, and Extra Trees. They find that random forests best fit the purpose of categorization, because random forests fix the overfitting problem of the decision tree model by adding randomness to the model (Ikudo et al. 2018). Similarly, Boselli et al. (2018) compare the classification results of multilingual online job ads using Support Vector Machines (including SVM Linear, SVM RBF Kernel, Random Forests), and Artificial Neural Networks (ANNs) and they draw the conclusion that SVM classifiers, will get more stable results of classification and are more suitable to high dimensional feature spaces (Boselli et al. 2018). Javed et al. (2016) use the combination of several machine learning models to build a semi-supervised classification model and they also make an improvement to the combined model system by building a cascade of hierarchical vertical classifiers to make the classification more precise and accurate (Javed et al., 2016). Interestingly, Schierholz and Schonlausome (2021) apply the same algorithms to survey responses where workers are asked to describe their occupations.

Although the mentioned machine learning models provide efficient categorization methods for jobs, one disadvantage of these models is that they do not provide explainable information of the model behavior, that is, they act like a black box. For better understanding of the labor demand and the occupational structure of the labor market, it would be essential to understand why these methods classify jobs into one or different categories. Topic modeling has the promise to solve this problem: it extracts the underlying latent topics that appear in job advertisements, provides a list of keywords for each topic that can be understood as sets of required skills, and allows us to use the topics to classify the jobs. For example, Mauro et al., (2016) use Latent Dirichlet Allocation (LDA), one of the most popular topic modeling algorithms, to categorize skills in about 2,700 online job posts into nine topics, and then they assign jobs into these nine topics according to the distribution of keywords in each job description. Colace et al. (2019) contrasts LDA with unsupervised classification methods. After applying both approaches on a dataset of European job descriptions, the authors find that LDA is preferrable as it could provide much more detailed information on the classification of jobs. In addition, Djumalieva et al. (2018) highlighted another advantage of LDA for job classification as it allows a job to be categorized into more than one topic and thus help better capture the feature of a job.

## 2.2 Economic literature

Relying on the above-mentioned literature in text analytics, computer science and data science, labor economists have applied the mentioned techniques to answer important questions of labor economics. One of the first published papers is Spitz-Oener (2006) who studies the long-term changes in skill requirements in West Germany and their relationship with computerization and skill-biased technological change. She categorizes occupational skill requirements into five categories, including nonroutine analytical, nonroutine interactive, routine cognitive, routine manual and nonroutine manual skills. Her findings show that the emphasis on skill requirements had shifted from cognitive and manual routine tasks to nonroutine analytical and interactive activities during the period from 1979 to 2006, and especially so in computerized occupations. The author concludes that the increased requirements of complex skills accounted for 36% of the change in the demand for higher-level education in employment.

Deming and Kahn (2018), analyze a dataset containing 45 million ads in the U.S. for the years 2010– 2015 provided by Burning Glass Technologies and categorize the skill requirements in job advertisements into 10 groups. The authors explore the correlation between each skill demand in the 10-skills categorization and the relative wages of the posted positions. Their results show that the firm-level wage variation can be explained by the differences in skill demand among firms. Meanwhile, the authors also find that within the 10 skill categories, social and cognitive skills have the strongest explanatory power of wage variation.

Hershbein and Kahn (2018) use the same dataset as Deming and Kahn (2018), as well as an additional dataset covering the posted job vacancies from 2007 in the U.S. labor market. Comparing the two datasets, they study how skill requirements changed over the Great Recession, focusing in particular on the cognitive and computer skills requirements. Their results indicate a possible structural change in skill demand because of the Great Recession as after the economic decline, the skill requirements increased in the metropolitan areas which were hit harder by the recession. The authors conclude that companies used the Great Recession to upgrade their technologies and skill demands, which conclusion was also supported by data on increased IT capital investments.

Ziegler (2021) studies the impact of skill requirements on wages using a vacancy database from the public employment administration (AMS) in Austria. The author finds that after accounting for occupation and firm fixed effects, there exists a positive relationship between the number of skill requirements mentioned in the job advertisements and the wages of the positions. While the association between managerial and analytical skills and wages are robust and large, the explanatory power of most soft skills regarding wages are relatively small.

Azar et al. (2020) studied labor market concentration using a dataset collected by Burning Glass Technologies which includes all job vacancies in the U.S. in 2016. They constructed a labor market concentration index applying the Herfindahl-Hirschman index to the vacancy data. The authors find a negative relationship between wages and labor market concentration. Meanwhile, their analysis also shows that there is no correlation between labor market concentration and the required skill level.

Atalay et al. (2020) construct a novel dataset with job advertisements published in three major metropolitan newspapers: the Boston Globe, the New York Times, and The Wall Street Journal and study the evolution of skill demand in the U.S. between 1950 and 2000. The authors use two established categorizations of skill requirements, the 5-skill categories by Spitz-Oener (2006) and the 10-skill categories by Deming and Kahn (2018). Concurring with the findings of prior studies, their results also show a long-term shift in skill demand from routine cognitive and manual skills to nonroutine interactive and analytic skills but the changes they detect are larger than what prior research identified.

Deming and Noray (2020) examine the impact of changing skill requirements on the wage dynamics of college graduates with a dataset provided by Burning Glass Technologies including all job vacancies posted online between 2007 and 2019. The authors reach three conclusions. Firstly, college graduates with more career-oriented majors, e.g. computer science, engineering and business, have higher starting wages. Secondly, in fast-changing fields, e.g. STEM, the returns to experiences are lower, which results in the flatter wage growth for the graduates working in these fields. Thirdly, college graduates with good learning abilities exit fast-changing STEM fields earlier.

Based on a dataset from a job-posting platform, CareerBuilder.com, Marinescu and Wolthoff (2020) build a model to detect the influence of firm and worker heterogeneity on the relationship between the wages of the job vacancies and the number of applicants for these positions. The authors find that the unexpected result that positions offering higher wages attract fewer applicants, which is due to job title heterogeneity. Within the same job title, higher wages attract more applicants.

# 3. Data description and preparation

## 3.1. Summary statistics

Our study uses job advertisement data from the Reed website, a comprehensive job advertisement platform owned by UK's largest online employment agency. The initial dataset for this project consists of 3,577,509 online job advertisements posted in the UK in 2018 and contains 13 variables. For this analysis, we make use of the following variables: 1) job title or position name; 2) job category, showing which industry the job belongs to; 3) company name, indicating which company has posted the job ad; 4) location, specifying which county in the UK the workplace is located in; 5) posting date of the job ad; 6) job description and requirements, elaborating on the expected qualifications, responsibility of the employee, welfare allowances or other important information the applicants should know about the position; 7) job type, indicating whether it is a full-time job (that is, more than 35 work hours per week) or part-time job (that is, less than 35 work hours per week), and whether it is a permanent or a temporary job.

Our data cleaning process included the deletion of missing, duplicated and non-conforming data, for example, we removed job ads with workplaces outside of the UK. We also dropped job ads with wages below or above the bottom or top 0.5 percentile of the wage data, respectively, which represented implausibly low or high wages. In the end, 1,158,926 lines of data were used for our subsequent analysis.

Figure 1 shows the distribution of jobs across the 35 industries in the dataset, using the categorization available on the job board website. The largest fraction of jobs represents business and IT-related jobs, education, engineering, construction, health and social care, and administrative jobs. In terms of location, the dataset covers almost all counties and regions of the UK, ranging from England, Scotland, to Wales and Northern Ireland. As a result, the dataset after preprocessing was representative enough to reflect the overall situation of the job market in the UK in 2018.

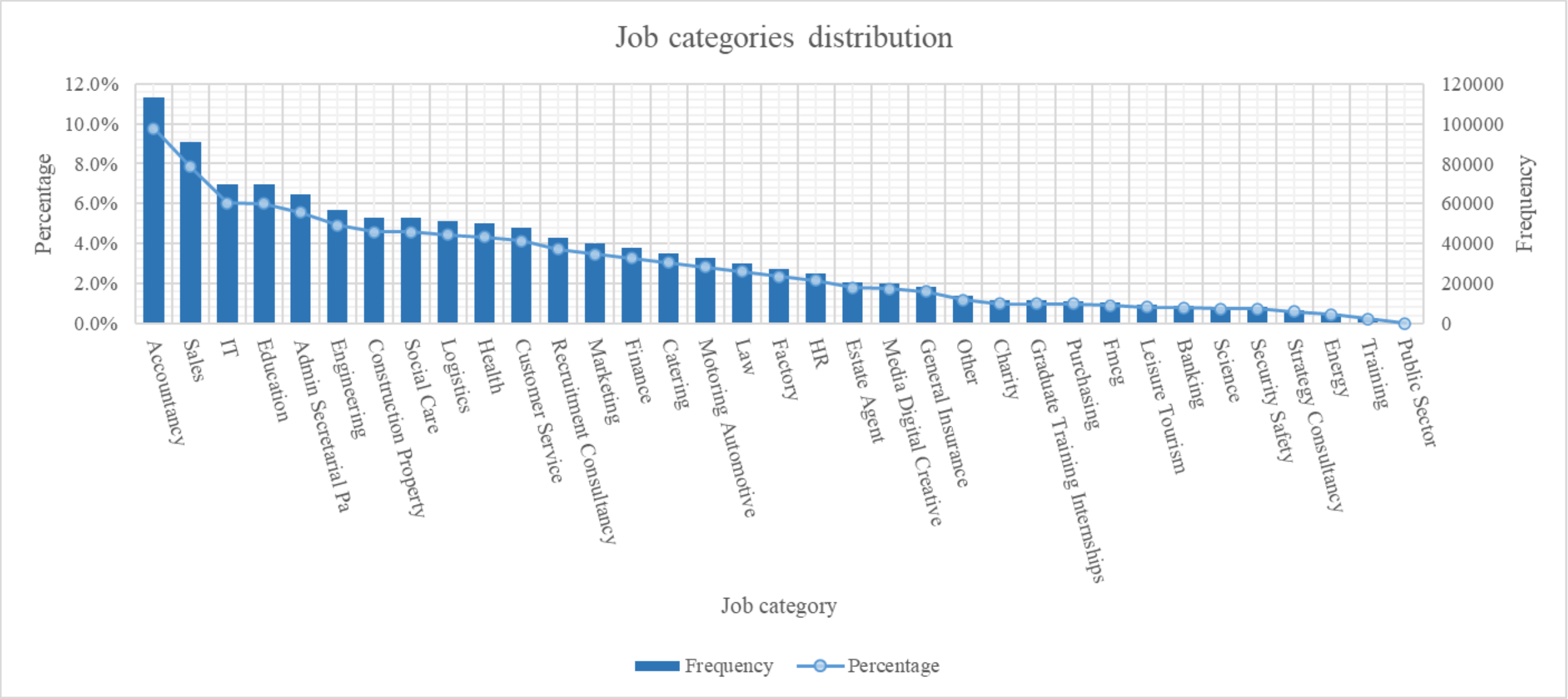

**Figure 1. Job categories distribution in the UK in 2018.**

Turning to the wage data, Figure 2 depicts the wage distribution. The average wage in the dataset is about 31 thousand pounds per annum and the maximum wage is about 10 times the minimum wage. As illustrated in Figure 2, most offered wages are concentrated at the lower end below the mean, and within the range from 10 thousand to 60 thousand pounds per annum. To get an idea of the wage variation across and within job categories, we calculated the mean wage and the Gini coefficient for each of the 35 job categories in the data. Gini coefficient is defined to quantitatively measure the overall wage variation based on the wage distribution within the specific job category. The higher the coefficient, the more unequal the distribution tends to be. Figure 2 shows the overall Gini coefficient for the whole dataset is 0.264, which is slightly lower than the value reported by the Office for National Statistics in the UK, 0.347 (Clark 2021). We conjecture that our dataset oversamples jobs with higher education requirements, and thus wages, that are more likely to be advertised online. We therefore have less jobs in the low wage range than the nationally representative data provided by the Office for National Statistics. The Gini coefficient also varies among the job categories ranging from 0.13 to 0.3, a reasonable rage. Our main goal is to explain the observed wage variation in the sample based on the skill requirements in the job advertisements, and compare how different skill categorization methods are able to explain these wage variations.

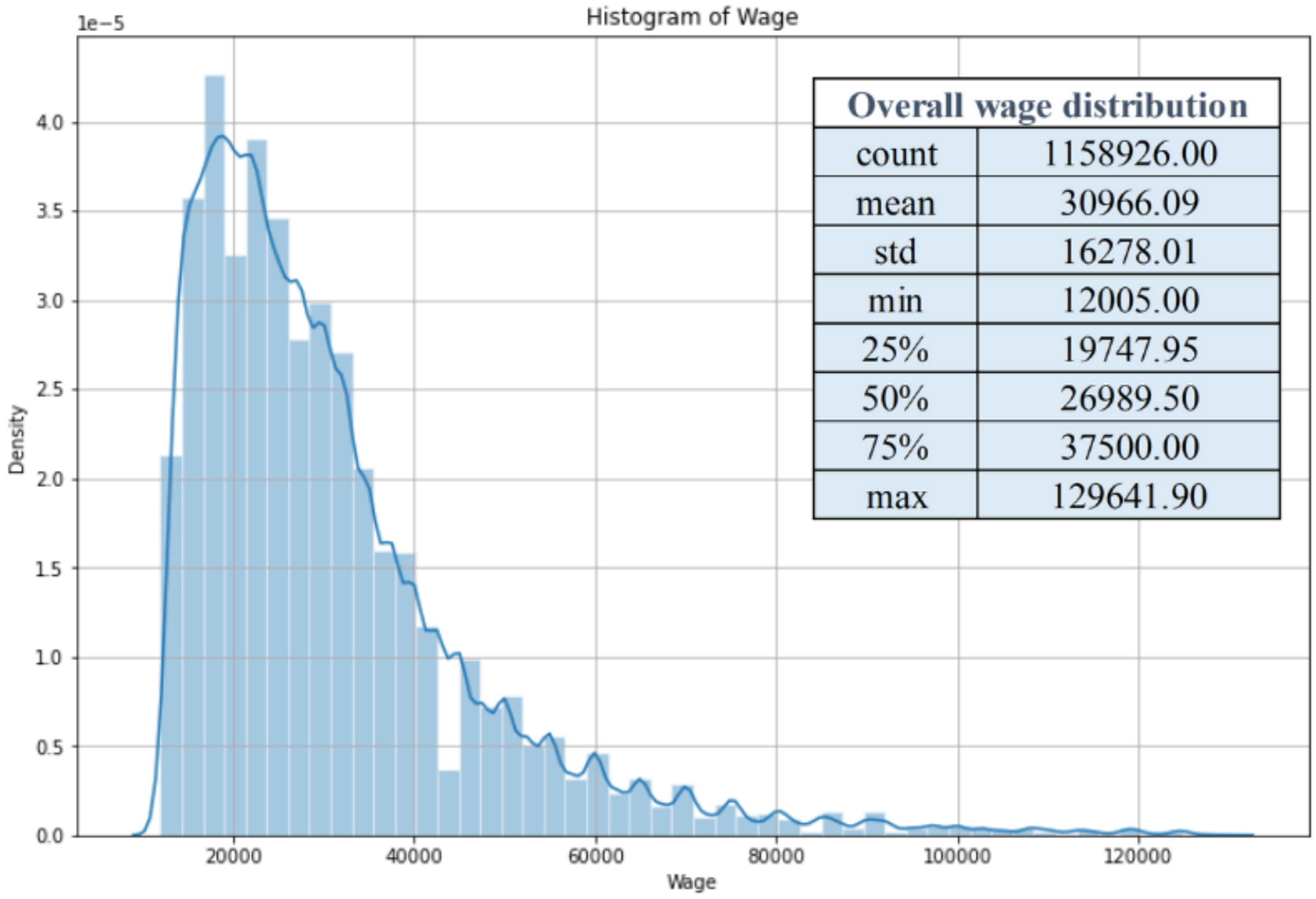

**Figure 2. Overall descriptive data of the job wage distribution in the UK in 2018.**

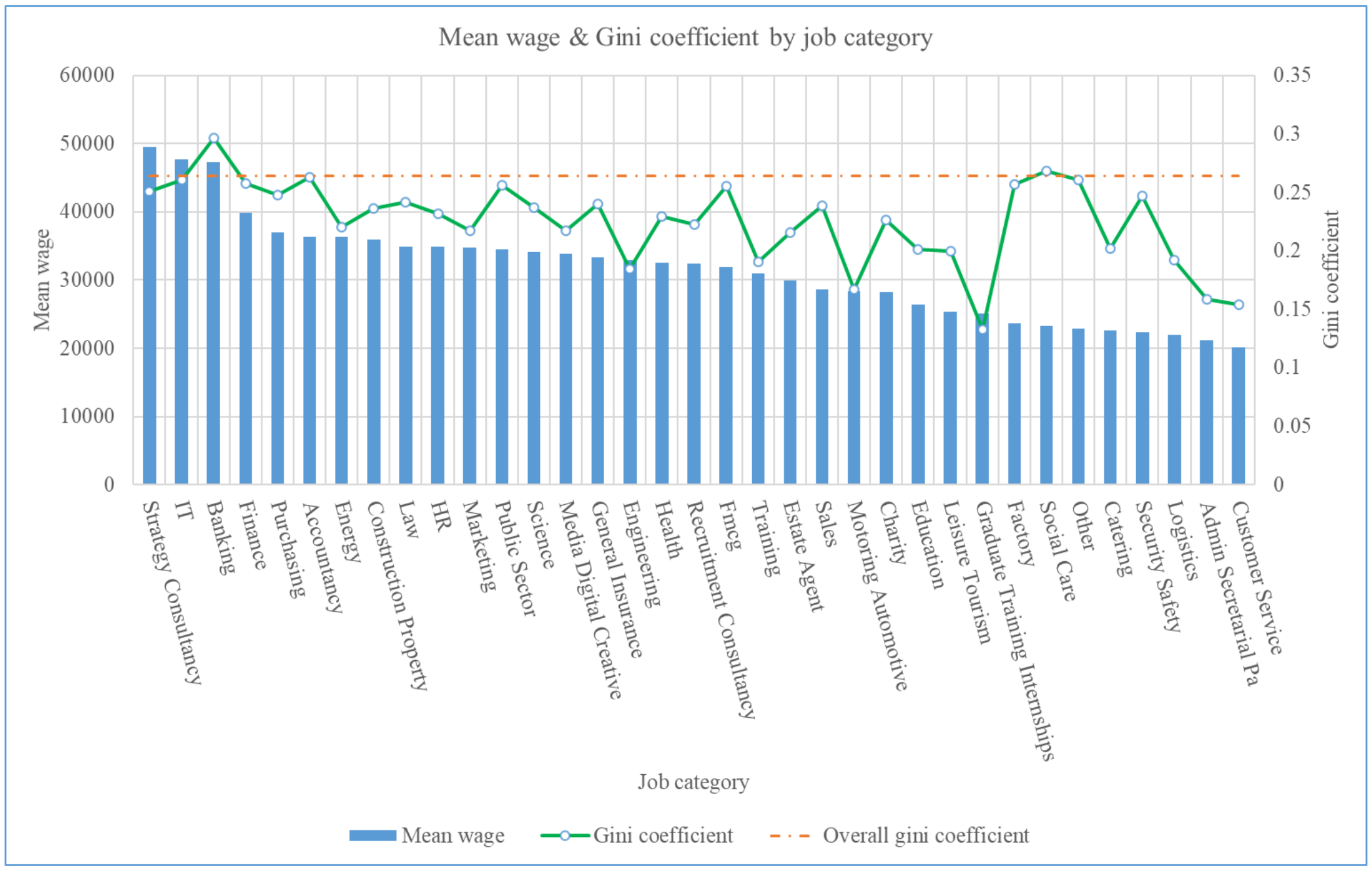

**Figure 3. The mean wage for each job category in descending order, their corresponding Gini coefficients, and the overall wage Gini coefficient for the whole dataset in 2018.**

## 3.2 Text preparation

We cleaned the job descriptions from each job advertisement by using standard methods in text analytics. The process of text normalization is shown in Figure 4. We convert all words to lower case, remove HTML tags, extra lines, numbers, accented chars, special characters and the extra white spaces. We remove all stopwords using the list of keywords provided in the NLTK package of Python for the English language. We lemmatize the text which means the removal of inflectional endings and keeping the base or dictionary form of each word. Text normalization helps us to extract meaningful information more efficiently, and avoid the repetition of words with different forms but the same meaning, which is conducive to improving the accuracy of the methods we use for further text analytics.

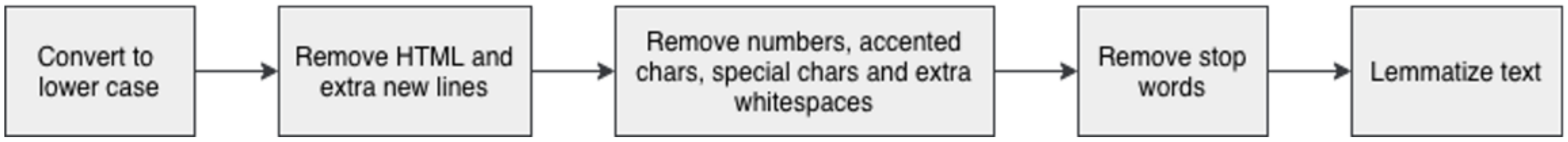

**Lemmatize*: remove inflectional endings only and return the base or dictionary form of the word

**Figure 4. Flow of string normalization**

# 4. Construction of skills

We construct the skill requirements embedded in the job description of ads using four different methods. The first three methods are based on word counting using predefined keywords of predefined skill types that were established by the labor economic literature. The last method apply unsupervised topic modeling techniques whereby both the skill types and the corresponding keywords are determined by the algorithms from the bottom up. Importantly, none of the methods use the wage data which we will use to compare the different methods.

## 4.1 Skill sets from the literature

Spitz-Oener (2006) proposed the assignment of skills into 5 categories while Deming and Kahn (2018) suggested a broader classification of 10 skill categories. Their categorization of job skills are widely applied models in the field of labor economics for identifying occupational skill requirements. They are applicable across a wide range of jobs because they represent general skills. As shown in Table 1 and Table 2, each of these classifications defines a list of 5 or 10 skills and provides the corresponding words and phrases to describe a series of activities that fall into each category. The 5-skill classification of Spitz-Oener (2006) mainly focuses on finding the worker's job duties or tasks, depicted by the activities that employees have to perform in the workplace that fall into these 5 broad classes. The 5 skill categories include the following: nonroutine analytic, nonroutine interactive, routine cognitive, routine manual and nonroutine manual.

The 10-skill classification of Deming and Kahn (2018) aimed to find important skills or traits that would explain pay differentials across labor markets and performance differentials across firms. The 10 skills groups include the following: cognitive, social, character, writing, customer service, project management, people management, financial, general and specific computer skills. They are derived from the aforementioned 5-skill classification and other previous literature, which results in a modified version of the 5-skill classification that is applicable to a variety of jobs. For example, the "cognitive" skill is designed to match the "nonroutine analytic" skills. It extends the noncognitive or "soft" skills into "social" and "character", explaining the importance of personality traits (Deming and Kahn, 2018).

When we apply these skill classifications to the data, we expanded the keywords relative to the lists provided in Table 1 and Table 2 since the job descriptions might not use the exact same words as shown in the tables. Therefore, we added synonyms, different versions and substitutes of each keyword. The words were expanded using Python packages and then edited manually. The complete lists of keywords used in the data analysis are relegated to the Appendix (see Table A1 and A2).

Based on the list of keywords, we construct a *skill intensity index* for each skill in each job advertisement. We count the number of keywords of a given skill category that appear at least once in a given job ad's job description and divide it by the total number of keywords of that skill category. For example, if a skill category has 10 keywords out of which 3 appears at least once in a job description, we say that the intensity of that skill for this job ad is 3/10=30%. In this way, we compute the skill intensity for all listed skills and for all job advertisements. Note that one job may require multiple skills in this way. We have also experimented with a binary measure, which assigns 1 to a skill if at least one of the keywords appears in the job description, and 0 otherwise. The results of that analysis are very similar to the one that we present in the paper.

Figure 5 shows the average skill intensity index for each of the 5 skill categories from Spitz-Oener (2006). We can see that the least demanded skill category is routine manual skills, while the most demanded is non-routine interactive skills. The average skill intensity of the other three skills is at about 80%. Figure 6 depicts a similar graph based on the 10 skill categories of Deming and Kahn (2018). We can see that the most demanded skill category is consumer service at 75%, while other popular skills include financial, project management and social skills. The least common skill category is specific computer skills at 2% of the jobs.

**Table 1. Assignment of Activities (Spitz-Oener 2006)**

| Classification | Tasks |
|---|---|
| Nonroutine analytic | Researching, analyzing, evaluating, and planning, making plans/constructions, designing, sketching, working out rules/prescriptions, and using and interpreting rules |
| Nonroutine interactive | Negotiating, lobbying, coordinating, organizing, teaching or training, selling, buying, advising customers, advertising, entertaining or presenting, and employing or managing personnel |
| Routine cognitive | Calculating, bookkeeping, correcting texts/data, and measuring length/weight/temperature |
| Routine manual | Operating or controlling machines and equipping machines |
| Nonroutine manual | Repairing or renovating houses/apartments/machines/vehicles, restoring art/monuments, and serving or accommodating |

**Table 2. Description of Job Skills (Deming and Kahn 2018)**

| **Job Skills** | **Keywords and Phrases** |
|---|---|
| Cognitive | Problem-solving, research, analytical, critical thinking, math, statistics |
| Social | Communication, teamwork, collaboration, negotiation, presentation |
| Character | Organized, detail-oriented, multitasking, time management, meeting deadlines, energetic |
| Writing | Writing |
| Customer service | Customer, sales, client, patient |
| Project management | Project management |
| People management | Supervisory, leadership, management (not project), mentoring, staff |
| Financial | Budgeting, accounting, finance, cost |
| Computer (general) | Computer, spreadsheets, common software (e.g., Microsoft Excel, PowerPoint) |
| Computer (Specific) | Programming language or specialized software (e.g., Java, SQL, Python) |

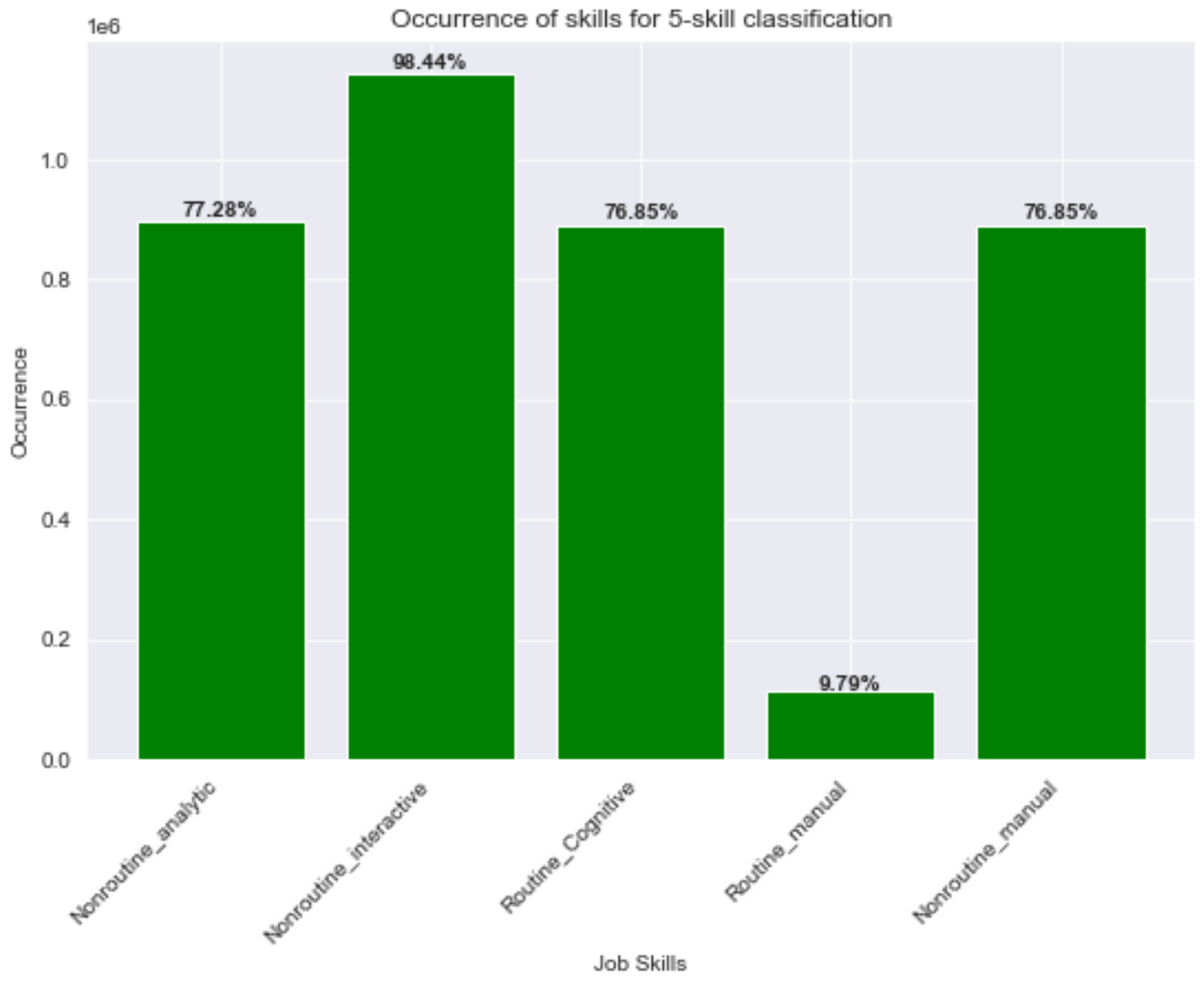

**Figure 5: Fraction of jobs requiring each skill based on the 5-skills categorization**

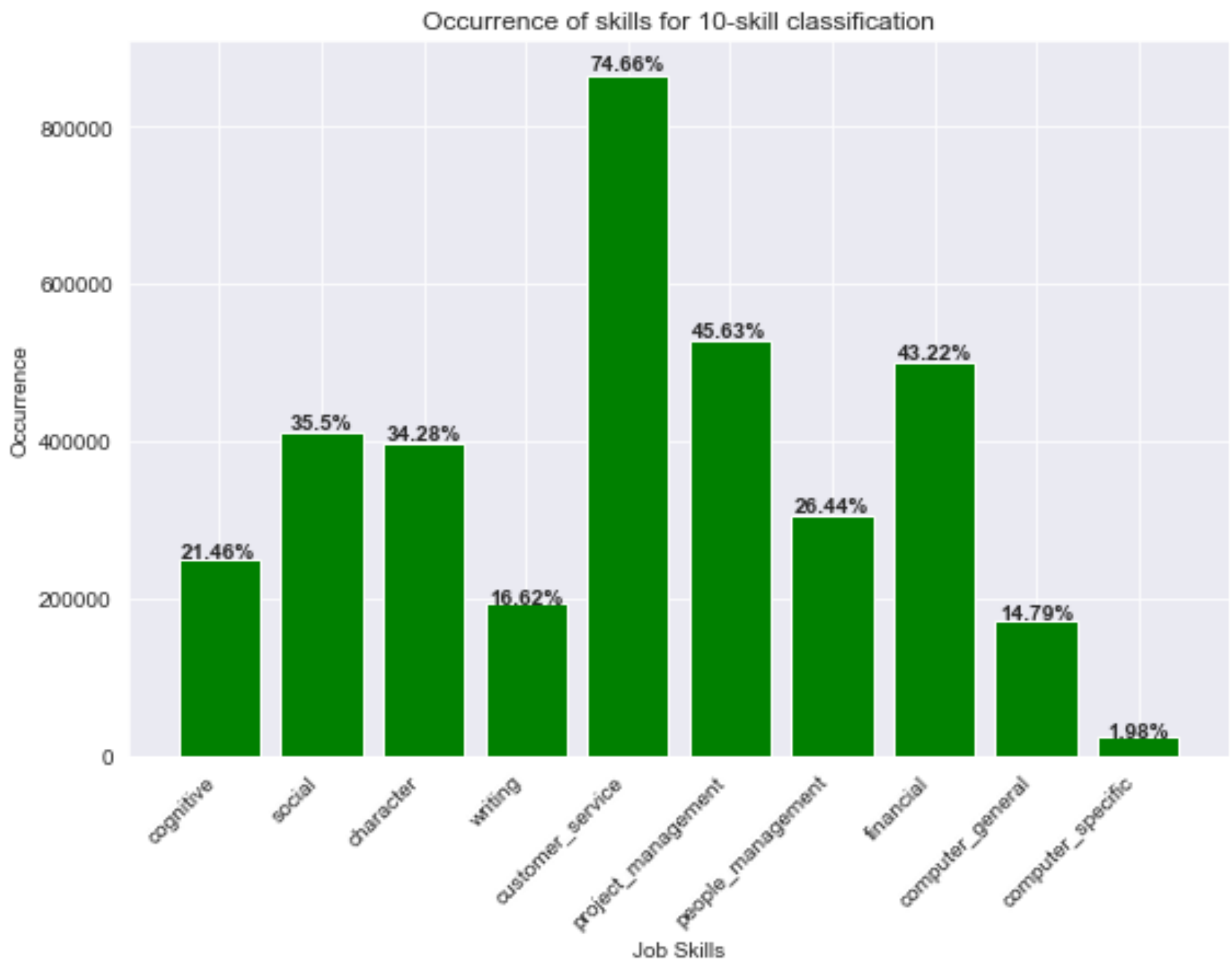

**Figure 6: Fraction of jobs requiring each skill based on the 10-skills categorization**

## 4.2 European Dictionary of Skills and Competencies (DISCO)

Our third skill categorization method is based on the European Dictionary of Skills and Competences, which we refer to as DISCO (DISCO, n.d.) in the sequel. It is a multilingual, peer-reviewed thesaurus used to classify, describe and translate skills and competencies, which has been incorporated in the European Classification of Skills, Competences, Occupations, and Qualifications. DISCO divides skills into 8 non-domain-specific categories (artistic, basic action verbs, computer, driving licenses, languages, managerial & organizational, materials tools, products & software, personal, social, communication skills & competencies), which capture more general skill keywords that may be broadly required by different jobs in the labor market. In addition, DISCO incorporates 25 domain-specific skill categories, which give detailed professional skill keywords that are only required by jobs in a particular domain. The domain-specific categories incorporate skills in the fields of law, health, business, architecture, etc., which could directly distinguish between industries. For instance, the skill keywords for the architecture domain includes light design, traffic planning, urban redevelopment, etc., which are required by different kinds of occupations within this field respectively.

Skill keywords are presented in a hierarchical way in the DISCO dictionary and we use the keywords in the last layer of each job category to conduct the classification of skills in the job advertisements. Specifically for the non-domain-specific categories, we excluded the basic action verbs category, which incorporates too many general verbs to make a distinction between job ads and we kept the other categories. The list of keywords for the remaining categories can be found in the Appendix (see Table A3).

Regarding the 8 non-domain specific skills in DISCO, we use the same method as in Section 4.1. for the 5- and 10-skills classification to establish the skill requirements of each job advertisement. Specifically, we create a skill intensity for each job and skill type which measures the fraction of the skill's keywords that appear in the job description. Figure 7 shows the average skill intensity for each of the 9 non-domain specific skills. The most commonly required skills include social and communication skills, managerial and organizational skills and computer skills.

As for the domain specific skills in DISCO, the first task is to assign each job advertisement to one of the 25 domains of DISCO. We perform this task by using the Word2vec word embedding model. Specifically, there are two steps given as:

1. **Word embedding:** We converted the unique keywords of skills in DISCO and the job descriptions in our dataset to a vector representation using the ***Word2vec word embeddings model*,** which was trained using a combination of the two datasets. After that, we calculated the **simple averaging** of all keywords word embedding vectors for each job description and skill, then the average value would represent that job/skill category.

2. **Similarity measurement:** For each job, we measured the similarity between job description and each of the 25 DISCO domain-specific skill category vectors, using ***cosine similarity***. Then, the element-wise means of resulting vectors of cosine similarities were calculated and each job was assigned to the category with the highest average similarity (Djumalieva et al., 2018).

Once we matched each job advertisement to a domain in DISCO, we calculate the skill intensity of each job's domain by computing the fraction of domain-specific keywords that the job advertisement contains. Figure 8 shows the average skill intensity score for each of the 25 domains in DISCO.

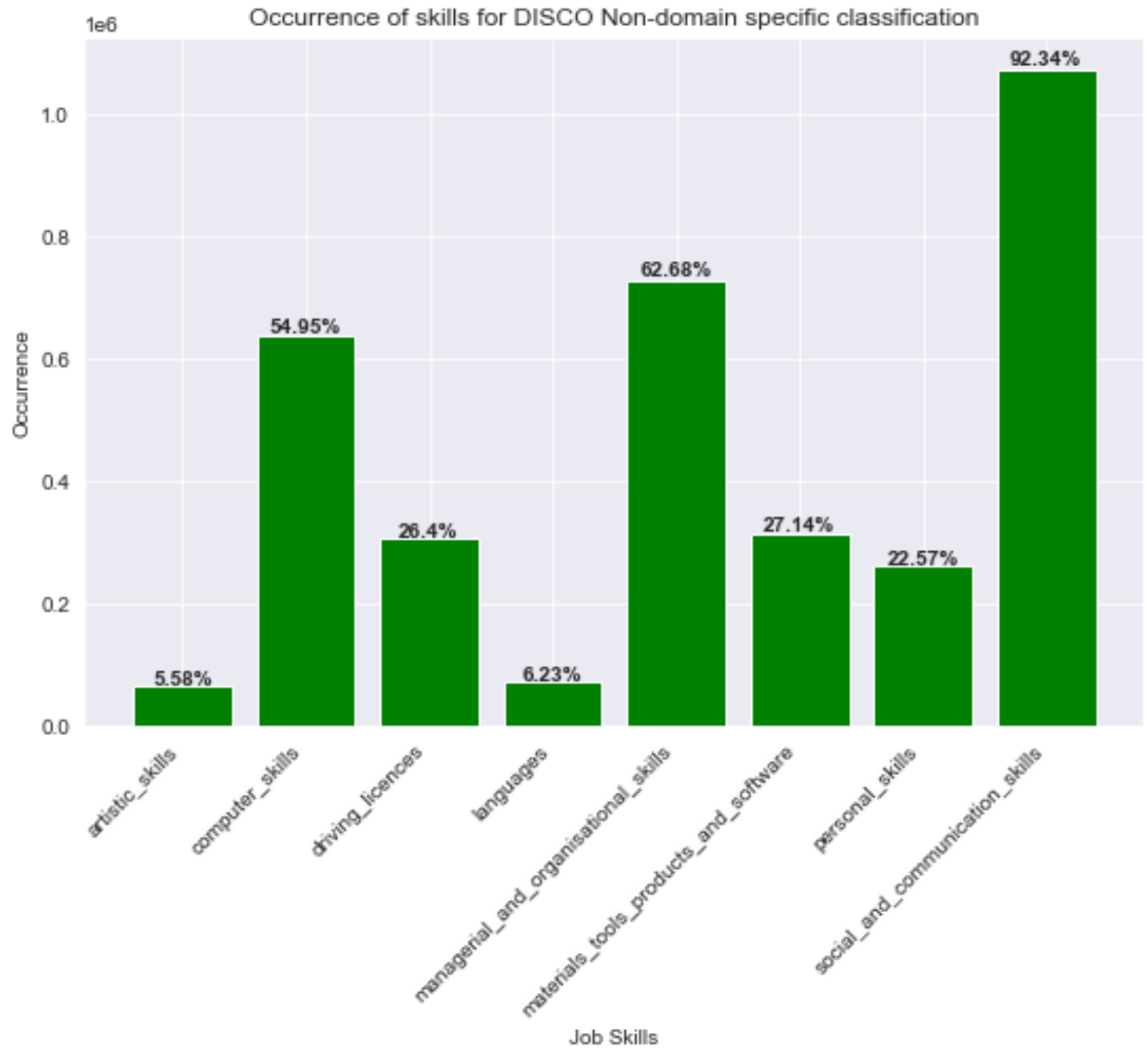

**Figure 7. Fraction of jobs requiring each skill based on the DISCO non-domain specific skills**

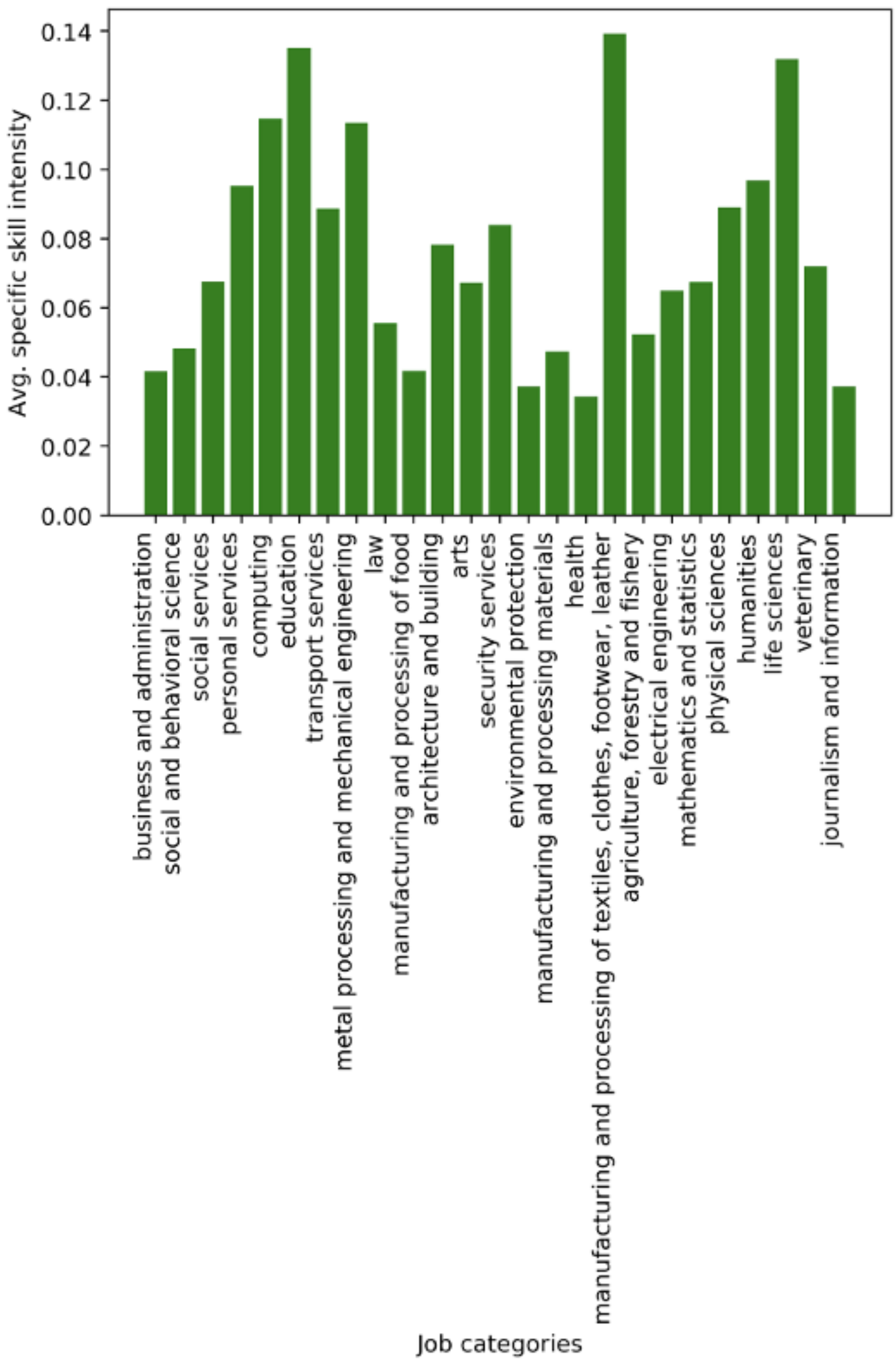

**Figure 8. Skill intensity of DISCO domain-specific skill categories**

## 4.3 Topic modeling: Latent Dirichlet Allocation (LDA)

Our last job categorization method is based on unsupervised topic modeling. Topic modeling is a type of statistical modeling for discovering the abstract topics that occur in a collection of documents (Nikolenko et al., 2016). In our case this means to return in an unsupervised way different skill categories and associated keywords that appear in the job advertisements. Latent Dirichlet Allocation (LDA) is an example of a topic model which is developed using Variational Exception Maximization (VEM) algorithm for obtaining the maximum likelihood estimate from the whole corpus of text. The LDA model follows the concept that each document can be described by the probabilistic distribution of latent topics and each topic can be described by the probabilistic distribution of words. Each topic is, in turn, modeled as an infinite mixture over an underlying set of topic probabilities (Blei et al., 2003). LDA provides keywords for each topic and the corresponding positive weights per topic for each document in the corpus, where the value of weights is proportional to the representativeness of the topic in the document.

In our context, we adapted LDA to categorize all the job descriptions into different topics, which represented skill requirements by a collection of keywords that belonged to each topic. Two 100,000 randomly generated sample data from the whole dataset was trained to jointly find the optimal number of topics based on the coherence score. The coherence score in topic modeling measures how similar the top 30 keywords with the highest probability of belonging to that particular topic are to each other. Therefore, the higher the score is, the better results we could get. We utilized the CV coherence score in our study, which created content vectors of words using their co-occurrences and, after that, calculated the score using normalized pointwise mutual information (NPMI) and the cosine similarity (Zvornicanin, 2021).

Before training, we filtered out words that occurred in less than 20 documents, or more than 60% of the documents to improve the model efficiency. The optimal topic number we get, which is 24, was used to train the LDA model on the whole dataset, which returned 24 topics each with keywords and corresponding weights. To quantify the skill requirements of each job, we calculated the distribution of the 24 topics for each job advertisement based on the probability that each topic could be assigned to this job. Each job advertisement has 24 probability scores corresponding to the 24 generated topics, and these scores add up to one. The topic with the highest probability is defined as the *dominant topic* of that job advertisement. Probabilities on the 24 topics for each job will be utilized in the wage regression analysis in Section 5. Table A4 in the Appendix shows the keywords of each generated topic, and distribution of job advertisements over the 24 topics where each job is assigned to its dominant topic. We can observe that the generated topics correspond to meaningful job skills. For example, topic 1 corresponds to financial and accounting-related skills, topic 2 gathers education-related keywords, topic 3 corresponds to health services, topic 4 is associated to HR-related skills, etc. However, some of the topics are hard to interpret, especially that they seem to overlap with other topics. For example, both topic 22 and 17 correspond to sales-related activities. There are also several topics that tough upon different aspects of management. We thus conclude that the interpretation of the skills identified by the LDA model might be challenging at times.

# 5. Results: Comparison of methods by wage regressions

In summary, this study has explored the skill requirements in job advertisements in the UK with various skill-categorization methods. We in total apply three predefined skill categories, specifically 5 skills, 10 skills and DISCO, and the LDA model to perform supervised and unsupervised text classification respectively.

In this section, we compare the different skill-categorization methods presented in Section 4 by the variation of wages they can explain based on wage regressions. The underlying idea is that firms reward skills by higher wages and they set the wages paid by each job based on the quantity and type of skills the employee needs to possess to perform well in the job. Our dataset contains the wages paid by each job and the skill requirements are embedded in the job descriptions. A better skill categorization method thus should be able to perform better at extracting the relevant skill requirements from the job description and the more accurate the established skill requirements are, the higher fraction of the wage variation they can explain. To obtain a measure of the wage variation explained by skill requirements, we run wage regressions and compute the *adjusted R2* from the regressions which also takes into account that the different regressions will have different numbers of explanatory variables.

To be more specific, we run wage regressions of the following type:

$$\log(w_i) = \beta_0 + \beta_1 s_i + \beta_2 x_i + \varepsilon_i$$

where $w_i$ is the wage of the job, $s_i$ is a vector of skill requirements of job $i$, $x_i$ is a vector of other job characteristics, $\varepsilon_i$ is the error term. $\beta_1$ is a vector of the skill premia associated to the skills contained in $x_i$. More specifically, we standardize the different skill intensity measures introduced in Section 4 to have zero mean and standard deviation one, and we use the standardized scores in the wage regression. The coefficient $\beta_1$ can thus be interpreted as the percentage increase in wage when the skill intensity increases by one standard deviation. We run three versions of this regression with varying controls present in $x_i$: dummy variables for job type, the month the job was posted, and job location. Our preferred regression contains all control variables and is presented as Model 3 in the following regression tables.

Table 4 shows the regression results with the skill intensity measures based on the 5-skills classification from Spitz-Oener (2006) as discussed in Section 4.1. Results show that the intensity of all five skills is significantly correlated with the wage variable but their effects on the expected salary are not all positive. While the coefficients of nonroutine interactive, nonroutine manual and routine manual skills are negative, the positive coefficients of nonroutine analytic and routine cognitive skills indicate a positive impact of these two skill requirements on the wage. Nonroutine analytic skill owns the largest coefficient among the five skills. These findings are robust to the inclusion of additional control variables (see Model 2 and 3). While all skill intensity variables are significant, the adjusted R2 of our preferred regression in Model 3 is quite small at 0.208.

**Table 4: Wage regression based on the 5-skills classification from Spitz-Oener (2006)**

| | Model1 | Model2 | Model3 |
|---|---|---|---|
| nonroutine analytic | 0.138*** | 0.133*** | 0.124*** |
| | (0.000) | (0.000) | (0.000) |
| nonroutine interactive | -0.032*** | -0.029*** | -0.030*** |
| | (0.000) | (0.000) | (0.000) |
| routine cognitive | 0.016*** | 0.015*** | 0.015*** |
| | (0.000) | (0.000) | (0.000) |
| routine manual | -0.022*** | -0.019*** | -0.015*** |
| | (0.000) | (0.000) | (0.000) |
| nonroutine manual | -0.027*** | -0.024*** | -0.018*** |
| | (0.000) | (0.000) | (0.000) |
| Constant | 10.231*** | 10.231*** | 10.231*** |
| | (0.000) | (0.000) | (0.000) |
| Job type FE | No | Yes | Yes |
| Posting month FE | No | Yes | Yes |
| County FE | No | No | Yes |
| R2adj | 0.099 | 0.130 | 0.208 |
| R2adj within | . | 0.094 | 0.088 |
| Number of observations | 1,158,926 | 1,158,926 | 1,158,926 |

Similarly, Table 5 summarizes the regression results using the intensity of 10-skills based on the classification from Deming and Kahn (2018). The intensity of all 10 skills are significantly correlated with the wage variable. The coefficients of character-related skills, writing, customer service and general computer skills are negative and the coefficients of the other six skills are positive. Moreover, the project management skill, with the highest positive coefficient. We also obtain that the adjusted R2 of our preferred regression Model 3 is 0.227, a very similar value to the one we obtain for the 5-skills categorization. The wage variation explained by the skill intensities are thus not very different between the 5-skills and the 10-skills categorizations.

**Table 5: Wage regression based on the 10-skills classification from Deming and Kahn (2018)**

| | Model1 | Model2 | Model3 |
|---|---|---|---|
| cognitive | 0.028*** | 0.027*** | 0.018*** |
| | (0.000) | (0.000) | (0.000) |
| social | 0.008*** | 0.008*** | 0.003*** |
| | (0.000) | (0.000) | (0.000) |
| character | -0.018*** | -0.019*** | -0.019*** |
| | (0.000) | (0.000) | (0.000) |
| writing | -0.018*** | -0.019*** | -0.019*** |
| | (0.000) | (0.000) | (0.000) |
| customer_service | -0.003*** | -0.007*** | -0.005*** |
| | (0.000) | (0.000) | (0.000) |
| project_management | 0.119*** | 0.114*** | 0.106*** |
| | (0.000) | (0.000) | (0.000) |
| people_management | 0.002*** | 0.003*** | 0.005*** |
| | (0.000) | (0.000) | (0.000) |
| financial | 0.072*** | 0.068*** | 0.062*** |
| | (0.000) | (0.000) | (0.000) |
| computer_general | -0.047*** | -0.044*** | -0.040*** |
| | (0.000) | (0.000) | (0.000) |
| computer_specific | 0.047*** | 0.046*** | 0.040*** |
| | (0.000) | (0.000) | (0.000) |
| Constant | 10.231*** | 10.231*** | 10.231*** |
| | (0.000) | (0.000) | (0.000) |
| Job type FE | No | Yes | Yes |
| Posting month FE | No | Yes | Yes |
| County FE | No | No | Yes |
| R2adj | 0.130 | 0.159 | 0.227 |
| R2adj within | . | 0.124 | 0.110 |
| Number of observations | 1,158,926.000 | 1,158,926.000 | 1,158,926.000 |

In addition, Table 6 shows the regression results between skill intensities and wages using the skill intensities constructed from DISCO, as introduced in Section 4.2. This includes the 8 non-domain specific skills and the intensity of domain-specific skills that are based on the matching of each job ad to a domain. The coefficients of all skills are statistically significant. The coefficients of language, social communication, material and personal skills are negative in all regressions, but the other coefficients are positive. Managerial and organization skill possesses the largest positive coefficient. The R2 of our most preferred Model 3 stands at 0.2, again a similar value compared to the previous two classification methods. We can thus conclude that the wage variation explained by the three top-down skill categorization methods, that use expert opinion on skill keywords to detect skills in job advertisements, stands at about 20%.

Lastly, Table 7 outlines the wage regressions using the skill intensities that were constructed from the 24 topics based on unsupervised LDA topic modeling algorithm. We note that not all coefficients of the 24 topic skill intensities are statistically significant and among the significant ones, the sign of the coefficients varies between negative and positive. We find that the R2 of the regressions are much higher than the R2s from our previous regressions based on supervised, top-down skill categorization. In particular, for our preferred Model 3 the R2 is as high as 0.483. This is especially remarkable considering that the topic modeling algorithm did not use the wage information as input in any way.

One might consider that it is not completely fair to compare our 24-topic LDA model to the other supervised skill categorization methods because the latter have a much smaller number of skill categories. In Table 8, we analyze the robustness of our findings by considering a 10-skills version of the LDA model. We retrain the LDA topic model by setting the number of topics to 10, in order to make the number of skills comparable to ones used in our previous analyses. We again find that the wage regression based on the LDA model has a much higher R2 than the regressions based on the previous skill categorization methods. The adjusted R2 of the wage regression based on the 10-topic LDA model is 0.455.

In sum, we find that all skill categorization methods point to skills that significantly influence the wages, and can differentiate between skills that have positive or negative impact on the wages. More importantly, we obtain that the unsupervised LDA topic modeling algorithm is the most successful at recovering skills from the job descriptions in the sense that it can explain the largest fraction of wage variation among all categorization methods considered. While LDA skills explain the highest fraction of wage variation, not all skills/topics identified by this method are easy to interpret based on the suggested keywords. One advantage of the supervised, top-down methods remains in the easiness of interpretation of the different skill types.

**Table 6: Wage regression based on the DISCO skill classification**

| | Model1 | Model2 | Model3 |
|---|---|---|---|
| Domain specific skill intensity | -0.047*** | 0.086*** | 0.055*** |
| | (0.012) | (0.012) | (0.011) |
| *Non-domain specific skill intensities* | | | |
| Artistic | 0.019*** | 0.017*** | 0.015*** |
| | (0.000) | (0.000) | (0.000) |
| Computer | 0.035*** | 0.032*** | 0.029*** |
| | (0.000) | (0.000) | (0.000) |
| Driving | 0.007*** | 0.006*** | 0.009*** |
| | (0.000) | (0.000) | (0.000) |
| Languages | -0.021*** | -0.022*** | -0.028*** |
| | (0.000) | (0.000) | (0.000) |
| Managerial organization | 0.128*** | 0.120*** | 0.110*** |
| | (0.000) | (0.000) | (0.000) |
| materials | -0.018*** | -0.015*** | -0.010*** |
| | (0.000) | (0.000) | (0.000) |
| Social and communication | -0.023*** | -0.023*** | -0.023*** |
| | (0.000) | (0.000) | (0.000) |
| Personal skills | -0.032*** | -0.027*** | -0.023*** |
| | (0.000) | (0.000) | (0.000) |
| Constant | 10.236*** | 10.224*** | 10.226*** |
| | (0.001) | (0.001) | (0.001) |
| Job type FE | No | Yes | Yes |
| Posting month FE | No | Yes | Yes |
| County FE | No | No | Yes |
| R2adj | 0.091 | 0.120 | 0.200 |
| R2adj within | . | 0.084 | 0.078 |
| Number of observations | 1,158,926.000 | 1,158,926.000 | 1,158,926.000 |

**Table 7: Wage regression based on skill classification constructed by LDA topic modelling (24 topics)**

| | Model1 | Model2 | Model3 |
|---|---|---|---|
| topic1prob | 0.022*** | 0.022*** | 0.021*** |
| | (0.005) | (0.005) | (0.005) |
| topic2prob | -0.024*** | -0.023*** | -0.026*** |
| | (0.005) | (0.005) | (0.005) |
| topic3prob | -0.027*** | -0.025*** | -0.018*** |
| | (0.005) | (0.005) | (0.004) |
| topic4prob | -0.010*** | -0.011*** | -0.006* |
| | (0.004) | (0.003) | (0.003) |
| topic5prob | -0.031*** | -0.032*** | -0.028*** |
| | (0.003) | (0.003) | (0.003) |
| topic6prob | 0.132*** | 0.130*** | 0.127*** |
| | (0.005) | (0.005) | (0.005) |
| topic7prob | 0.045*** | 0.044*** | 0.047*** |
| | (0.004) | (0.004) | (0.004) |
| topic8prob | -0.037*** | -0.030*** | -0.025*** |
| | (0.002) | (0.002) | (0.002) |
| topic9prob | 0.025*** | 0.024*** | 0.022*** |
| | (0.003) | (0.003) | (0.003) |
| topic10prob | 0.067*** | 0.066*** | 0.067*** |
| | (0.005) | (0.005) | (0.005) |
| topic11prob | 0.028*** | 0.026*** | 0.027*** |
| | (0.005) | (0.005) | (0.005) |
| topic12prob | 0.018*** | 0.017*** | 0.008** |
| | (0.003) | (0.003) | (0.003) |
| topic13prob | 0.009*** | 0.008*** | 0.011*** |
| | (0.003) | (0.003) | (0.003) |
| topic14prob | 0.021*** | 0.020*** | 0.010*** |
| | (0.004) | (0.004) | (0.004) |
| topic15prob | 0.003 | 0.002 | 0.006*** |
| | (0.002) | (0.002) | (0.002) |
| topic16prob | -0.001 | 0.004 | 0.004 |
| | (0.004) | (0.004) | (0.003) |
| topic17prob | -0.068*** | -0.068*** | -0.059*** |
| | (0.003) | (0.003) | (0.003) |
| topic18prob | -0.000 | -0.001 | -0.001 |
| | (0.002) | (0.002) | (0.002) |
| topic19prob | -0.058*** | -0.058*** | -0.053*** |
| | (0.003) | (0.003) | (0.003) |
| topic20prob | -0.069*** | -0.066*** | -0.060*** |
| | (0.004) | (0.004) | (0.004) |
| topic21prob | -0.094*** | -0.094*** | -0.096*** |
| | (0.005) | (0.005) | (0.004) |
| topic22prob | -0.018*** | -0.012*** | -0.010*** |
| | (0.001) | (0.001) | (0.001) |
| topic23prob | 0.109*** | 0.107*** | 0.102*** |
| | (0.004) | (0.004) | (0.004) |
| topic24prob | -0.042*** | -0.039*** | -0.039*** |
| | (0.004) | (0.004) | (0.004) |
| Constant | 10.231*** | 10.231*** | 10.231*** |
| | (0.000) | (0.000) | (0.000) |
| Job type FE | No | Yes | Yes |
| Posting month FE | No | Yes | Yes |
| County FE | No | No | Yes |
| R2adj | 0.441 | 0.448 | 0.483 |
| R2adj within | . | 0.425 | 0.404 |
| Number of observations | 1,158,926.000 | 1,158,926.000 | 1,158,926.000 |

**Table 8: Wage regression based on skill classification constructed by LDA topic modelling (10 topics)**

| | Model1 | Model2 | Model3 |
|---|---|---|---|
| topic1prob | 0.199*** | 0.196*** | 0.181*** |
| | (0.010) | (0.009) | (0.009) |
| topic2prob | -0.016 | -0.015 | -0.021** |
| | (0.010) | (0.010) | (0.009) |
| topic3prob | -0.034*** | -0.024** | -0.022** |
| | (0.011) | (0.011) | (0.011) |
| topic4prob | -0.097*** | -0.094*** | -0.087*** |
| | (0.012) | (0.012) | (0.011) |
| topic5prob | 0.097*** | 0.094*** | 0.093*** |
| | (0.009) | (0.009) | (0.008) |
| topic6prob | 0.019*** | 0.018*** | 0.005 |
| | (0.006) | (0.006) | (0.006) |
| topic7prob | 0.034*** | 0.032*** | 0.033*** |
| | (0.011) | (0.011) | (0.011) |
| topic8prob | -0.089*** | -0.089*** | -0.090*** |
| | (0.009) | (0.009) | (0.009) |
| topic9prob | -0.046*** | -0.041*** | -0.051*** |
| | (0.009) | (0.009) | (0.009) |
| topic10prob | 0.016 | 0.015 | 0.014 |
| | (0.011) | (0.011) | (0.010) |
| Constant | 10.231*** | 10.231*** | 10.231*** |
| | (0.000) | (0.000) | (0.000) |
| Job type FE | No | Yes | Yes |
| Posting month FE | No | Yes | Yes |
| County FE | No | No | Yes |
| R2adj | 0.408 | 0.418 | 0.455 |
| R2adj within | . | 0.394 | 0.372 |
| Number of observations | 1,158,926.000 | 1,158,926.000 | 1,158,926.000 |

## 6. Conclusion

Our paper has compared several methods used in the literature to extract skill requirements from the text of job advertisements. These methods used in different disciplines. Most economic papers prefer to use a top-down approach that defines skills and a set of corresponding skills based on expert-opinion and searches for those keywords in the text of job advertisements. The data and computer science literature, in contrast, uses automated bottom-up approaches such as topic modeling. This approach uses unsupervised algorithms to identify topic or skills in the job advertisements, where the topics/skills and the corresponding keywords are endogenously determined by the algorithm.

We performed three top-down approaches, two based on the skill classification of specific economic papers (Deming and Kahn, 2018; Spitz-Oener, 2006), and one based on the European Dictionary of Skills and Competences (DISCO). We also applied the LDA topic modeling on our dataset U.K. job advertisement. We used each method to understand the prevalence of skill requirements in the data. Then, we run wage regression to estimate the returns or discounts that are attached to each identified skills. Ultimately, we have compared the different skill extraction methods based on the wage variation that the extracted skill can explain.

We find that the LDA model performs the best in terms as it explains about 45% of the observed wage variation in the labor market. In contrast, each of the three top-down methods can explain only about 20% of the wage variation. The performance of the LDA model is remarkable since we did not use the wage information in the LDA algorithm to identify the required skills stated in the job advertisements. One drawback of the LDA models is, however, that the identified topics are sometimes hard to interpret as a specific skill requirement, which may be a disadvantage in labor economic applications. Relatedly, some of the topics are hard to distinguish from each other as similar or the same keywords appear in several topics. The top-down approaches do not have these drawbacks; however, they cannot fully recover the skills that are relevant for the wages in the labor market. This is indicated by the relatively low wage variation they can explain.

Future work should concern about improving the interpretability of the LDA results, and compare the LDA model and the top-down approaches to further topic modeling techniques.

# Appendix

**Table A1.Expanded Assignment of Activities (Spitz-Oener 2006)**

| Classification | Tasks |
|---|---|
| Nonroutine analytic | research, research, search, explore, analyze, analyse, study, examine, canvass, canvas, analyze, analyse, break down, dissect, take apart, analyze, analyse, analyze, analyse, psychoanalyze, psychoanalyse, analysis, analysis, analytic thinking, analysis, analysis, analysis, psychoanalysis, analysis, depth psychology, measure, evaluate, valuate, assess, appraise, value, evaluate, pass judgment, judge, planning, planning, planning, preparation, provision, plan, be after, plan, plan, project, contrive, design, design, plan, construct, build, make, manufacture, fabricate, construct, construct, construct, construct, reconstruct, construct, retrace, construction, building, construction, grammatical construction, expression, construction, mental synthesis, structure, construction, construction, construction, twist, construction, building, design, designing, plan, project, contrive, design, design, design, design, plan, design, design, design, designing, scheming, sketch, chalk out, sketch, outline, adumbrate, prescription, prescription drug, prescription, prescription medicine, ethical drug, prescription, prescription, interpret, construe, see, rede, interpret, interpret, render, represent, interpret, translate, interpret, render, understand, read, interpret, translate |
| Nonroutine interactive | negociate, negotiate, talk terms, negotiate, negociate, lobby, buttonhole, organize, organise, coordinate, coordinate, coordinate, align, ordinate, coordinate, coordinating, coordinative, form, organize, organise, organize, organise, mastermind, engineer, direct, organize, organise, orchestrate, organize, organise, coordinate, organize, organise, prepare, devise, get up, machinate, unionize, unionise, organize, organise, teaching, instruction, pedagogy, teaching, precept, commandment, education, instruction, teaching, pedagogy, didactics, educational activity, teach, learn, instruct, teach, training, preparation, grooming, education, training, breeding, train, develop, prepare, educate, train, prepare, discipline, train, check, condition, prepare, groom, train, educate, school, train, cultivate, civilize, civilise, aim, take, train, take aim, direct, coach, train, train, train, train, rail, trail, train, selling, merchandising, marketing, sell, sell, sell, deal, sell, trade, sell, sell, sell, betray, sell, buying, purchasing, buy, purchase, bribe, corrupt, buy, grease one's palms, buy, buy, buy, ad, advertisement, advertizement, advertising, advertizing, advert, advertising, publicizing, advertise, publicize, advertize, publicise, advertise, advertize, promote, push, entertain, entertain, think of, toy with, flirt with, think about, harbor, harbour, hold, entertain, nurse, entertaining, show, demo, exhibit, present, demonstrate, present, represent, lay out, stage, present, represent, present, submit, present, pose, award, present, give, gift, present, deliver, present, introduce, present, acquaint, portray, present, confront, face, present, present, salute, present, use, utilize, utilise, apply, employ, hire, engage, employ, force, personnel, personnel department, personnel office, personnel, staff office |
| Routine cognitive | calculate, cipher, cypher, compute, work out, reckon, figure, calculate, estimate, reckon, count on, figure, forecast, account, calculate, forecast, calculate, calculate, aim, direct, count, bet, depend, look, calculate, reckon, calculating, calculative, conniving, scheming, shrewd, bookkeeping, clerking, correction, rectification, correction, fudge factor, correction, correction, chastening, chastisement, correction, discipline, correction, correction, measurement, measuring, measure, mensuration, measure, mensurate, measure out, quantify, measure, measure, measure, evaluate, valuate, assess, appraise, value |
| Routine manual | arming, armament, equipping, equip, fit, fit out, outfit, equip, equipment |
| Nonroutine manual | repair, mend, fix, bushel, doctor, furbish up, restore, touch on, compensate, recompense, repair, indemnify, repair, resort, rectify, remediate, remedy, repair, amend, animate, recreate, reanimate, revive, renovate, repair, quicken, vivify, revivify, renovate, restitute, refurbish, renovate, freshen up, animate, recreate, reanimate, revive, renovate, repair, quicken, vivify, |
| | revivify, renovation, redevelopment, overhaul, renovation, restoration, refurbishment, restore, reconstruct, regenerate, restore, rejuvenate, restore, restitute, repair, mend, fix, bushel, doctor, furbish up, restore, touch on, restore, reinstate, reestablish, helping, portion, serving, service, serving, service of process, serve, function, serve, serve, service, serve, serve, help, serve, serve up, dish out, dish up, dish, serve, serve, serve well, serve, do, serve, attend to, wait on, attend, assist, serve, process, swear out, suffice, do, answer, serve, serve, serve, service, serve |

**Table A2. Expanded Description of Job Skills (Deming and Kahn 2018)**

| **Job Skills** | **Keywords and Phrases** |
|---|---|
| Cognitive | research, inquiry, enquiry, research, research, research, search, explore, analytic, analytical, analytic, analytical, mathematics, math, maths, statistics, statistic |
| Social | communication, communicating, communication, communication, teamwork, collaboration, coaction, collaboration, collaborationism, quislingism, negotiation, dialogue, talks, negotiation, presentation, presentation, presentment, demonstration, presentation, presentation, display, presentation, presentation, introduction, intro, presentation |
| Character | form, organize, organise, organize, organise, mastermind, engineer, direct, organize, organise, orchestrate, organize, organise, coordinate, organize, organise, prepare, devise, get up, machinate, unionize, unionise, organize, organise, organized, organized, organized, organised, unionized, unionised, energetic, energetic, gumptious, industrious, up-and-coming |
| Writing | writing, authorship, composition, penning, writing, written material, piece of writing, writing, writing, writing, committal to writing, write, compose, pen, indite, write, publish, write, write, drop a line, write, compose, write, write, write, save, spell, write, write, write, compose, pen, indite, write, publish, write, write, drop a line, write, compose, write, write, write, save, spell, write, write |
| Customer service | customer, client, gross sales, gross revenue, sales, sale, sale, sale, cut-rate sale, sales event, sale, sale, sales agreement, client, customer, client, node, client, guest, patient, affected role, patient role, patient, patient |
| Project management | undertaking, project, task, labor, project, projection, project, stick out, protrude, jut out, jut, project, project, project, project, project, plan, project, contrive, design, project, propose, visualize, visualise, envision, project, fancy, see, figure, picture, image, project, cast, contrive, throw, project, send off, project, externalize, externalise |
| People management | supervisory, leadership, leading, leadership, leaders, leadership, leadership, mentor, staff, staff, staff, faculty, staff, staff, staff, stave, staff, staff |
| Financial | budget, accounting, accounting, accountancy, accounting, accounting, accounting system, method of accounting, account, accounting, account statement, account, account, calculate, report, describe, account, account, answer for, finance, finance, finance, finance, finance, cost, monetary value, price, cost, price, cost, toll, cost, be, cost |
| Computer (general) | computer, computing machine, computing device, data processor, electronic computer, information processing system, calculator, reckoner, figurer, estimator, computer, spreadsheet, excel, stand out, surpass |
| Computer (Specific) | Java, coffee, java, Java, python, python, Python |

**Table A3. Expanded DISCO Non-domain Specific Skills**

| Job Skills | Keywords and Phrases |
|---|---|
| Artistic skills | acting, aesthetics, sensitivity, creativity, imitate, musicality, musical, painting, drawing, trend, fashion-consciousness |
| Computer skills | software, application, software, basic internet, internet knowledge, internet, database, database management, database system, data, data protection, data security, e-mail, specific software, file, managing files, manage files, middleware, network management, network software, network, office, software, microsoft, software development, develop software, operating system, server software, server |
| Driving licences | drive, driving, driving licence, special vehicle, heavy goods vehicle, motorcycle, light motorcycle |
| Languages | afrikaans, albanian, arabic, basque, belorussian, bengali, bosnian, bulgarian, cantonese, catalan, chinese, croatian, czech, danish, dutch, english, estonian, farsi, finnish, flemish, french, gaelic, german, greek, hebrew, hindi, hungarian, icelandic, italian, japanese, korean, latvian, lithuanian, macedonian, malay, maltese, mandarin , mother tongue, native language, norwegian, polish, portuguese, rätoromance, romanian, russian, sardinian, serbian, slovak, slovenian, spanish, swedish, thai, turkish, ukrainian, urdu, vietnamese, foreign language, foreign languages |
| Managerial and Organisational skills | coordinate, organize oneself, organize, multitasking, multitask, decision-making, entrepreneurial, entrepreneurship, leadership, manage finance, managing finance, financial resources, finance, financial, managing material, manage material, resource, material resource, managing personnel, personnel resources, personnel, managing work, work activities, management techniques, management, monitoring, organizational, organizational performance, operative planning, plan, operative, risk-taking, self-sufficiency, set standard, setting, set target, setting standard, setting target, strategic planning, planning, supervision, supervise, time management |
| Materials Tools Products and Software | antique, chemist's, clocks, watches, jewellery, gold article, silver article, clothing, cnc machines, cnc, computer system, construction machine, construction material, data storage, sound, photo device, video device, camera, data storage, sound device, recorder, decorative material, material, do-it-yourself, drug, electrical product, electronic product, electrical goods, flooring material, flooring, food, footwear, shoe care, furniture, gardening supplies, gauge, gifts, glass, porcelain, household goods, household, ironmongery, tool, machinery, it products, leather goods, leather, lighting, metal, motor vehicle parts, accessories, musical instrument, network component, paints, varnishes, pets, pet food, pet accessories, photographic, film equipment, film device, video equipment, printing machinery, printing device, publishing products, sanitary equipment, bathroom accessories, sanitary, sports equipment, equipment, accessory, stationery supplies, office supplies, stone, synthetic material, teaching material, tobacco, toiletries, toiletry, toy, two-wheeled vehicle, wallpapers textiles, household textiles, textile, wood, wood articles, wooden articles |
| Personal skills | cognitive skill, cognitive, problem solving, solving, personal, personal skill, personal ability, physical attribute, physical ability, physical, solve, attribute, characteristic, problem |
| Social and Communication skills | verbal, verbal expression, expression, clear diction, distinct diction, diction, client support, communication, communicate, foreign languages, professional communication, written expression, written, questioning, question, establish, interview, comprehend, teach, teamwork, establishing contact, contract, contracting, fostering contact, contract |
| | establishing, contract fostering, intercultural, international, interviewing, listening, comprehension, listening comprehension, negotiation, oral comprehension, oral, participate, participation, discussion, discuss, meeting, presentation, rhetorical, service orientation, service, orientation, teaching, team-working, team, cooperate, cooperation, cooperating |

**Table A4. Distribution of 24 Generated Topics over Job Ads based on the LDA Model**

| Dominant Topic | Job Ad Count | % Job Ads | Keywords per Topic |
|---|---|---|---|
| Topic1 | 73084 | 6.31 | finance, account, financial, business, management, payroll, accountant, report, include, monthly, accounting, credit, company, payment, process, control, invoice, reconciliation, reporting, skill, qualified, client, within, base |
| Topic2 | 60718 | 5.24 | school, teacher, education, teach, student, teaching, year, support, look, primary, child, assistant, pupil, opportunity, staff, need, good, learn, position, september, excellent, secondary, start, within |
| Topic3 | 63404 | 5.47 | care, support, home, service, provide, nurse, hour, need, pay, worker, training, staff, include, nursing, resident, health, within, people, opportunity, healthcare, clinical, time, register, patient |
| Topic4 | 44959 | 3.88 | employment, please, recruitment, application, agency, candidate, within, opportunity, contact, business, vacancy, act, we, look, permanent, cv, position, client, successful, day, specialist, information, uk, due |
| Topic5 | 25220 | 2.18 | production, product, quality, food, manufacturing, kitchen, chef, hotel, manufacture, standard, within, high, machine, company, use, skill, environment, head, restaurant, engineering, include, process, produce, good |
| Topic6 | 101522 | 8.76 | management, business, ensure, project, manage, manager, process, support, skill, key, development, develop, ability, within, report, plan, provide, include, deliver, performance, delivery, lead, responsibility, change |
| Topic7 | 60593 | 5.23 | engineer, company, engineering, maintenance, project, design, technical, service, site, electrical, construction, client, look, mechanical, base, industry, candidate, equipment, building, opportunity, training, within, system, commercial |
| Topic8 | 9179 | 0.79 | scheme, holiday, pension, day, benefit, care, include, discount, opportunity, support, salary, employee, bonus, bank, training, provide, reward, offer, plus, career, join, people, full, free |
| Topic9 | 27485 | 2.37 | manager, store, retail, management, lead, manage, brand, business, opportunity, customer, high, assistant, service, deliver, within, look, develop, ensure, drive, great, leader, development, strong, excellent |
| Topic10 | 81346 | 7.02 | client, firm, opportunity, financial, legal, within, service, practice, look, tax, join, insurance, excellent, base, property, offer, commercial, provide, business, career, include, lead, please, support |
| Topic11 | 104418 | 9.01 | sale, business, client, company, new, opportunity, account, within, manager, development, target, lead, drive, market, relationship, look, executive, salary, base, successful, candidate, k, excellent, service |
| Topic12 | 24643 | 2.13 | recruitment, consultant, trainee, graduate, candidate, year, client, company, career, recruiter, sector, management, k, consultancy, london, training, business, interview, include, sale, look, office, salary, market |
| Topic13 | 23240 | 2.01 | vehicle, technician, automotive, repair, car, job, service, part, please, dealership, level, workshop, require, cv, client, within, look, manager, motor, contact, motortrade, full, uk, standard |
| Topic14 | 51131 | 4.41 | marketing, digital, medium, campaign, client, event, social, brand, content, communication, agency, skill, creative, strategy, manage, across, company, executive, design, include, manager, product, create, opportunity |
| Topic15 | 17074 | 1.47 | new, job, need, employment, recruitment, you, we, temporary, permanent, business, worker, supply, privacy, agency, look, position, limit, disclaimer, hay, hayscouk, click, accept, copy, interested |
| Topic16 | 46523 | 4.01 | service, support, people, social, post, application, provide, require, include, child, community, skill, information, need, job, assessment, within, development, please, council, qualification, staff, deliver, professional |
| Topic17 | 27833 | 2.4 | customer, service, provide, excellent, product, call, skill, advisor, ensure, centre, deliver, company, ability, sale, contact, environment, opportunity, time, communication, target, high, need, within, order |
| Topic18 | 21358 | 1.84 | hr, support, security, service, system, employee, line, provide, network, technical, officer, skill, training, infrastructure, knowledge, include, company, base, office, management, issue, across, business, engineer |
| Topic19 | 36313 | 3.13 | driver, ensure, require, safety, company, good, site, warehouse, duty, include, time, drive, transport, licence, clean, carry, delivery, class, stock, area, equipment, maintain, standard, hgv |
| Topic20 | 51059 | 4.41 | hour, per, week, day, pay, pm, start, time, shift, available, look, must, full, monday, client, please, require, rate, base, position, candidate, permanent, month, temporary |
| Topic21 | 98526 | 8.5 | skill, client, office, support, ability, require, excellent, ensure, include, duty, administration, administrator, communication, good, system, company, order, within, time, provide, maintain, use, able, management |
| Topic22 | 4405 | 0.38 | supplier, delivery, customer, day, procurement, deliver, job, supplychain, product, service, purchase, part, uk, application, you, include, logistic, royalmail, stock, drive, category, start, hour, new |
| Topic23 | 50220 | 4.33 | software, development, technology, developer, skill, datum, design, company, project, use, product, solution, technical, client, analyst, test, web, application, business, knowledge, lead, system, new, opportunity |
| Topic24 | 54673 | 4.72 | you, people, need, look, we, well, career, great, help, get, make, want, good, you, take, opportunity, one, also, support, training, every, new, like, day |